\documentclass[11pt, a4paper,hyper]{JHEP3}

\usepackage{amsmath}
\usepackage{amssymb}
\usepackage{graphicx}
\usepackage{slashed}
\usepackage{textcomp}

\def\tr{{\rm tr}}

\def\d{\delta}

\def\S{{\bf S}}

\def\CA{{\cal A}}

\def\CG{{\cal G}}
\def\CI{{\cal I}}
\def\CL{{\cal L}}
\def\CN{{\cal N}}
\def\CO{{\cal O}}

\def\CR{{\cal R}}
\def\CS{{\cal S}}

\def\b{\beta}
\def\g{\gamma}
\def\d{\delta}

\def\k{\kappa}

\def\m{\mu}
\def\n{\nu}

\def\r{\rho}

\def\s{\sigma}

\def\S{\Sigma}

\def\centeron#1#2{{\setbox0=\hbox{#1}\setbox1=\hbox{#2}\ifdim
   \wd1>\wd0\kern.48\wd1\kern-.48\wd0\fi
   \copy0\kern-.48\wd0\kern-.48\wd1\copy1\ifdim\wd0>\wd1
   \kern.48\wd0\kern-.48\wd1\fi}}

\newcommand{\beq}{\begin{equation}}
\newcommand{\eeq}{\end{equation}}
\newcommand{\bea}{\begin{eqnarray}}
\newcommand{\eea}{\end{eqnarray}}
\newcommand{\ba}{\begin{array}}
\newcommand{\ea}{\end{array}}

\newcommand{\p}{\partial}
\newcommand{\nn}{\nonumber}

\newcommand{\half}{\frac{1}{2}}

\newcommand{\bg}{{\bar{g}}}

\title{On Classical Equivalence Between Noncritical and Einstein  Gravity : The AdS/CFT Perspectives}

\author{Seungjoon Hyun$^a$, Wooje Jang$^a$, Jaehoon Jeong$^a$ and Sang-Heon Yi$^b$\\
$^a${\it\small{Department of Physics, College of Science, Yonsei University, Seoul 120-749, Korea}}\\
$^b${\it\small{Center for Quantum Spacetime, Sogang University, Seoul 121-741, Korea}}\\
E-mail: \email{sjhyun@yonsei.ac.kr},~\email{wooje@yonsei.ac.kr},~\email{j.jeong@yonsei.ac.kr},\\ \email{shyi@sogang.ac.kr}}

\abstract{ We find that noncritical gravity, a special class of higher derivative gravity,  is classically equivalent to Einstein gravity at the full nonlinear level. We obtain the viscosity-to-entropy ratio and the second order transport coefficients of the dual fluid of noncritical gravity to all orders in the coupling of higher derivative terms. We also compute the holographic entanglement entropy in the dual CFT of noncritical gravity.  All these results  confirm the nonlinear equivalence between noncritical gravity and Einstein gravity at the classical level.}
\keywords{Higher Curvature Gravity, Fluid-Gravity Correspondence, Holographic Entanglement Entropy }

\begin{document}
\maketitle
\section{Introduction}
Recently gravity with  higher derivative terms on Anti-de Sitter (AdS) spacetime has attracted much attention. Typically, higher derivative gravity theories contain ghost-like massive modes. An interesting observation in some class of higher derivative gravity on AdS spacetime  is that these ghost-like massive modes fall off more slowly than massless modes as they approach the boundary of AdS spacetime. Then we may consistently truncate ghost-like massive modes by imposing appropriate boundary conditions, leaving massless modes only.

The first example which incorporated this idea was provided in \cite{Maldacena:2011mk}, which is four-dimensional conformal gravity on AdS spacetime. The action is given by the conformally invariant Weyl-squared term only.  The quantum fluctuations of conformal gravity on AdS spacetime include massless and ghost-like massive modes. It was claimed that the ghosts can be removed by Neumann boundary conditions on the metric at the boundary and thus solutions of Einstein gravity can be chosen as a consistent subset among whole solutions of conformal gravity. Based on this observation, it was claimed that four-dimensional conformal gravity on AdS spacetime  may describe the same low energy physics as four-dimensional pure Einstein gravity.

Another interesting example  which incorporated this idea, so-called four-dimensional noncritical Einstein-Weyl gravity,  was proposed in \cite{Lu:2011ks} as an extension of critical gravity \cite{Lu:2011zk}. These theories
include Weyl-squared term with a specific value of coupling in addition to the usual Einstein-Hilbert and negative cosmological constant terms. They admit AdS spacetime as the vacuum solution.  The quantum fluctuations of the metric on this vacuum solution contain massless and ghost-like massive/logarithmic modes. These ghosts can be truncated by the boundary conditions as they fall off more slowly than massless modes toward the AdS boundary. In the case of critical gravity, the theory after the truncation might be trivial as  the excitation energy of massless modes is zero. In the case of  four-dimensional noncritical Einstein-Weyl gravity on (asymptotically) AdS spacetime, it was anticipated that the theory might be consistent after the truncation, which we call NEW gravity, with better UV behavior than usual Einstein gravity. 
 This was extended in \cite{Lu:2011mw} to $\CN =1$ NEW supergravity. 
 
 Though these studies are mostly based on linearized equations of motion, they suggest that  NEW gravity may be well-defined at low energy. Furthermore they indicate that the theory may describe essentially the same physics as Einstein gravity at the tree level.
Indeed, the classical equivalence between NEW gravity and Einstein gravity on AdS background at the full nonlinear level was established in \cite{Hyun:2011ej}. This was achieved by  showing that the effective Lagrangian of  NEW gravity becomes identical with the one of  Einstein gravity up to the rescaling of Newton's constant.  It was also shown in \cite{Hyun:2011ej} the classical equivalence between  ${\cal N}=1$ NEW supergravity and   ${\cal N}=1$ AdS supergravity.

From the perspectives of the AdS/CFT correspondence, the full classical equivalence between NEW gravity and Einstein gravity on AdS background  implies the equivalence between the corresponding dual conformal field theories(CFT) in the large $N$ limit. Indeed it was shown in \cite{Hyun:2011ej} that  the boundary action of NEW gravity after using  equations of motion is the same as the one of Einstein gravity up to the same overall rescaling of Newton's constant.
As a result, it was found that 
$n$-point correlation functions of energy-momentum tensor in the dual CFT of NEW gravity can be read off from those from Einstein gravity. It was also shown that  $n$-point correlation functions among energy-momentum tensors with two supercurrents in the dual CFT of ${\cal N}=1$  NEW supergravity is identical to those of the dual CFT of  ${\cal N}=1$ AdS supergravity.

In this paper we explore the implications of this classical equivalence further. Especially we are interested in the phenomena which reveal  nonlinear aspects of the equivalence through the AdS/CFT correspondence. 

First of all, we generalize the classical, but nonlinear equivalence between NEW gravity and Einstein gravity to arbitrary dimensions. In four dimensions, higher derivative terms in the action of critical gravity and NEW gravity are given by Weyl-squared term modulo Gauss-Bonnet terms. In higher dimensions, higher derivative terms of critical gravity are given by a specific combination of Ricci tensor-squared and Ricci scalar-squared terms with a special value of the coupling constant \cite{Deser:2011xc}. In this paper, we present the generalization of NEW gravity in higher dimensions, which we call noncritical gravity, whose action includes the same combination of the higher derivative terms with more general  value of the coupling constant.
Just like NEW gravity, noncritical gravity is defined on (asymptotically) AdS spacetime with the truncation of ghost-like massive modes of the metric by boundary conditions.
We establish  the classical equivalence between noncritical gravity and  Einstein gravity on AdS spacetime in arbitrary dimensions. All these are presented in section 2.

One of the most interesting arena which involves classical but nonlinear aspects on gravity is
the fluid/gravity correspondence. In section 3, we study the fluid/gravity correspondence for noncritical gravity. Among others, we obtain the linear and nonlinear transport coefficients for the dual fluid of noncritical gravity.   We find the exact viscosity-to-entropy ratio, all orders in higher derivative coupling, which saturates the KSS bound~\cite{Kovtun:2003wp}\cite{Kovtun:2004de} in accord with the previous perturbative result in the leading order of higher derivative coupling. 

Another interesting example which reveal nonpertubative aspects with respect to the higher derivative coupling is a
holographic entanglement entropy(HEE) of the dual CFT. In section 4, we propose two holographic methods which give  results consistent with expected nature of entanglement entropy. 
Once again we find  identical results for the HEE from noncritical gravity with those from  Einstein gravity. 

In section 5, we draw our conclusions.

\section{Noncritical Gravity in Various Dimensions}

In this section we generalize NEW gravity \cite{Hyun:2011ej} to arbitrary dimensions and show the classical, nonlinear equivalence between a class of the higher derivative gravity theories, so-called noncritical gravity,  and  Einstein gravity on AdS spacetime.

Let us consider the following
action 
\beq
S= \frac{\s}{2\k^2} \int d^D x \sqrt{-g} \left[ R +\frac{(D-1)(D-2)}{\ell^2} - \frac{1}{m^2(D-2)} \left(R^{MN}R_{MN} - \frac{D}{4(D-1)}R^2 \right)\right],\label{action1}
\eeq
where the overall sign $\s$ can take the value $\pm 1$ and   $\ell^2$ and $m^2$ take real values whose range will be determined later. For a special value for $m^2$, it becomes the action of critical gravity in arbitrary dimensions \cite{Deser:2011xc}.
The higher derivative terms can be rewritten as the combination of the Weyl-squared term and the Gauss-Bonnet term $E$ as
\bea
R^{MN}R_{MN} - \frac{D}{4(D-1)}R^2 =\frac{(D-2)}{{4(D-3)}}\left( C^{MNPQ}C_{MNPQ} - E \right)~.
\eea
The Gauss-Bonnet term $E= R^{MNPQ}R_{MNPQ}-4R^{MN}R_{MN} +R^2$ is total derivative in four dimensions and thus the above higher derivative terms in the action reduces to the Weyl-squared term in four dimensions. One interesting feature on this specific combination of curvature-squared terms is the absence of scalar modes.
Another  interesting feature  is that it satisfies the condition for higher curvature couplings  to be consistent with holographic $c$-theorem \cite{Myers:2010tj}.

The Euler-Lagrange equations of motion for the metric $g_{MN}$ are given by
\beq
G_{MN} - \frac{(D-1)(D-2)}{2\ell^2} g_{MN}+E_{MN}=0~, \label{4eom}
\eeq
where
\bea
G_{MN}&=& R_{MN}-\frac{1}{2}R g_{MN} \, ,\label{fulleom}\\
E_{MN}&=& \frac{1}{(D-2)m^2}\left[ -2(R_{MP}R_{N}^{~P} - \frac{1}{4} R^{PQ}R_{PQ} g_{MN} ) + \frac{D}{2(D-1)} R( R_{MN} - \frac{1}{4} R g_{MN}) \right. \nn \\
&& \left. - ( \nabla^2 R_{MN} + \half \nabla^2 R g_{MN} -2 \nabla_{P} \nabla_{(M} R_{N)}^{~P}) + \frac{D}{2(D-1)} ( g_{MN}\nabla^2 R - \nabla_{M}\nabla_{N} R)\right]~.\nn
\eea

It is convenient to introduce an auxiliary field $f_{MN}$ and rewrite the action as
\beq
S = \frac{\s}{2\k^2} \int d^D x \sqrt{-g} \left[ R +\frac{(D-1)(D-2)}{\ell^2} - \frac{1}{(D-2)} f^{MN} G_{MN} + \frac{m^2}{4(D-2)}  (f^{MN}f_{MN} - f^2)   \right]\,, \label{action11}
\eeq
which  was also used in \cite{arXiv:1102.4091} in the context of critical gravity.
The equation of motion for the auxiliary field $f_{MN}$ is given by
\beq
f_{MN} = \frac{2}{m^2} \left[ R_{MN} - \frac{1}{2(D-1)} R g_{MN} \right]\,,\qquad f = \frac{(D-2)}{m^2(D-1)} R\,.
\eeq
After plugging this equation of motion  back in the action, we recover the original action (\ref{action1}).
For appropriate ranges of $\ell^2$ and $m^2$, it admits AdS spacetime as vacuum solution,
\beq
\bar{R}_{MN} = -\frac{(D-1)}{L^2} \bg_{MN} \,, \qquad \bar{R} = -\frac{D(D-1)}{L^2}\,,\qquad \bar{G}_{MN} = \frac{(D-1)(D-2)}{2L^2} \bg_{MN}\,,
\eeq
and
\beq
\bar{f}_{MN} = -\frac{(D-2)}{m^2L^2}\bar{g}_{MN}\label{aux0} \,,
\eeq
where $\bar{g}_{MN}$ denotes the AdS metric and the AdS curvature radius $L$ is given by the relation,
\beq \frac{1}{\ell^2} = \frac{1}{L^2} \left[ 1 - \frac{(D-4)}{4m^2 L^2}\right] \,.  \label{NewCos}\eeq

The quantum fluctuations of the metric $g_{MN}$ around AdS spacetime, generically, consist of the massless and massive modes, one of which turns out to be ghost-like. A clever choice of the expansion is given by 
\bea
g_{MN} &=& \bar{g}_{MN}+ h_{MN}+\phi_{MN} \,,  \label{expand1}\\
f_{MN} &=& -\frac{(D-2)}{m^2 L^2} \Big( \bar{g}_{MN} +h_{MN}\Big)  -(D-2) \left(\frac{2L^2}{\ell^2} -1\right) \phi_{MN}  \,, \nn
\eea
which  leads  to the natural decoupling between massless and massive modes, denoted as $h_{MN}$ and $\phi_{MN}$, respectively.

Indeed, through these expansions, the quadratic Lagrangian ${\cal L}_2$,  modulo the total derivative terms,  is given by
\beq
\CL_{2} =-\frac{q}{2} h^{MN} \CG_{MN}(h) +\frac{q}{2} \phi^{MN} \CG_{MN}(\phi) + \s \frac{(D-2)}{4} m^{2} q^{2} ( \phi^{MN} \phi_{MN} -\phi^2 ) \,,
\eeq
where
\beq
q =  \s \left[1 - \frac{(D-2)}{2 m^2 L^2} \right]\,. \label{q}
\eeq
 Here ${\cal{G}}_{MN}$ denotes the linearized Einstein operator including the cosmological constant term,
\bea
 {\cal{G}}_{MN}(h)
 = R^{(1)}_{MN}(h) -\half \bar{g}_{MN}\bar{g}^{PQ} R^{(1)}_{PQ}(h)  +\frac{(D-1)}{L^2} h_{MN} - \frac{(D-1)}{2L^2} h \bar{g}_{MN}\,.
\eea
The linearized Ricci tensor,  $R_{MN}^{(1)}$,  is given by
\bea
R_{MN}^{(1)} (h)&=& \bar\nabla_P \bar\nabla_{(M} h^P_{N)} - \frac{1}{2}\bar\nabla_N \bar\nabla_M h - \frac{1}{2}\bar\nabla^2 h_{MN}\,,
\eea
where $\bar\nabla$ denotes a covariant derivative with respect to the background metric $\bar{g}_{MN}$.

Clearly, the massless and massive modes are decoupled in the above quadratic Lagrangian. The novel aspect of the decoupling  is that the bulk-to-boundary propagators in the AdS/CFT correspondence do not mix those two kinds of modes. One may also note that the signatures of the kinetic terms of the massless and massive modes are opposite, signaling one of them is ghost-like. By taking $q$ positive, the massless gravitons, $h_{MN}$,  remain physical, while the massive modes, $ \phi_{MN}$,  become ghost-like.
The corresponding linearized equations of motion are given by
\bea
  {\cal{G}}_{MN}(h) &=& 0\,,   \label{leom}
\\
  {\cal{G}}_{MN}(\phi) &+& \s \frac{(D-2)}{2} m^{2} q\left( \phi_{MN} - \phi \bar{g}_{MN} \right) =0 \,.\nn
 \eea

If we choose the
transverse traceless gauge, which is consistent with the equations of motion, as
\bea
\bar \nabla^{M} h_{MN} = 0 \,,\qquad \bg^{MN} h_{MN} = 0\,,\qquad \bar \nabla^{M} \phi_{MN} = 0 \,,\qquad \bg^{MN} \phi_{MN} = 0\,,
\eea
the linearized equation of motion for a massless graviton $h_{MN}$ reduces to
\bea
\left( \bar\nabla^2 + \frac{2}{L^2}\right)  h_{MN}=0\,,\label{eom1}
\eea
and the one for a ghost-like massive graviton $\phi_{MN}$, with mass $M$, becomes
\bea
 \left( \bar\nabla^2  + \frac{2}{L^2} - M^2 \right) \phi_{MN}=0\,, \qquad M^2=(D-2)\s qm^2\,.\label{eom2}
\eea

AdS spacetime has the time-like boundary on which the boundary conditions should be specified. A salient feature for the case when $M^2<0$  or $m^2<\frac{D-2}{2L^2}$  is that ghost-like massive modes, $ \phi_{MN}$,  fall off more slowly than massless modes, $h_{MN}$, as they approach  the boundary. Therefore, in this case, these ghosts can be consistently truncated by imposing appropriate boundary conditions. The BF bound\cite{Breitenlohner:1982jf} which comes from the tachyon-free condition is given by  $M^2 > -\frac{ (D-1)^2}{4L^2 }$, which is $m^2>\frac{D^2-6D+7}{4(D-2)L^2}$. Therefore we consider the action with the coupling range
\beq
\frac{D^2-6D+7}{4(D-2)L^2}< m^2<\frac{D-2}{2L^2}~.
\eeq
One may note that $m^2$ is always positive for $D\geq 5$ in contrast to the case $D=4$ where negative $m^2$ is allowed\footnote{  In three dimensions, analogous study has been carried out in 
\cite{Bergshoeff:2009hq}.}. Since $q$ should be positive to have physical massless modes, the overall sign of the action, $\s$, should be negative for $D\geq 5$ while  in four dimensions $\s$ can take both signs.
If $m^2$ saturates the upper bound, $m^2=\frac{D-2}{2L^2}$, the massive modes turn into the logarithmic ones, and the corresponding theory is called critical gravity.

Now comes the key point. As alluded earlier, ghost-like massive modes, $\phi_{MN}$, are decoupled from massless modes at the quadratic Lagrangian and therefore do not mix in the bulk-to-boundary propagator. Once  ghosts, $\phi_{MN}$, are truncated at the boundary, they remain suppressed  even deep in the bulk  at the tree level. After the truncation of ghosts, we have a linearized relation (\ref{expand1}) in which $f_{MN} $ is proportional to $g_{MN}$. Without ghosts,  this linearized relation can be consistently lifted to the full non-linear level as \cite{Hyun:2011ej}
\beq
f_{MN} = -\frac{D-2}{m^2L^2}g_{MN}\label{aux1} \,.
\eeq
This means that  solutions of Einstein gravity can be chosen as a consistent subset among whole solutions of our class of higher derivative gravity. We call this higher derivative gravity with the above consistent truncations of ghosts as noncritical gravity.

By
plugging this  in the action (\ref{action11}), we obtain
the effective action:
\beq
S_{eff} = \frac{q}{2\k^2} \int d^D x \sqrt{-g}\left[R + \frac{(D-1)(D-2)}{L^2} \right]~,
\label{action2}\eeq
which is nothing but the ordinary Einstein action with Newton's constant $\frac{\kappa^2}{q}$ and the cosmological constant $-\frac{(D-1)(D-2)}{2L^2} $. In particular
this strongly indicates that the dual conformal field theory of noncritical gravity would be identical to the one dual to Einstein gravity in the large $N$ limit.

Let us turn into boundary actions and show that similar arguments can be applied. The boundary action, which we use in the AdS/CFT correspondence, consists of the generalized Gibbons-Hawking terms and boundary terms from the on-shell bulk action. 
We introduce  the bulk metric in the ADM decomposed form as
\beq
ds^2 = N^2 dr^2 + \g_{\m\n} (dx^\m + N^\m dr)(dx^\n + N^\n dr)\,, \label{ADM}
\eeq
and define  the boundary fields $\hat{f}^{\m\n} $ by
\beq
\hat{f}^{\m\n} = f^{\m\n} + 2 h^{(\m}N^{\n)} + sN^\m N^\n\,, \quad  \hat{f} =
\gamma_{\m\n}\hat{f}^{\m\n}\,, \label{hatfuc} \eeq
from the  bulk auxiliary fields $f^{MN}$,
\beq
f^{MN} = \left(\begin{array}{cc}s & h^\n\nn\\
               h^\m & f^{\m\n} \end{array}  \right)\,.\eeq

Let us consider  the generalized Gibbons-Hawking terms \cite{Hohm:2010jc} at the boundary  which is given by
\beq
S_{GGH} = \frac{\s}{2\k^2} \int_{\partial M} d^{D-1}x \sqrt{-\g} \left[ -2 K + \frac{1}{(D-2)} \bigg(\hat{f}^{\m\n} K_{\m\n} - \hat{f} K \bigg) \right]\,, \label{baction1}
\eeq
where the extrinsic curvature $K_{\m\n}$ on the boundary surface  is defined by
\beq
K_{\m\n} = - \frac{1}{2N} (\p_r \g_{\m\n} - \nabla_{\m} N_\n - \nabla_{\n} N_\m )\,.
\eeq
After turning off  ghost-like massive modes of the boundary fields we can use the similar arguments for the boundary fields $ \g_{\m\n},\hat{f}_{\mu\nu} $ as for the bulk fields $ g_{MN}, f_{MN}$ and give the same relation,
\bea
\hat{f}_{\mu\nu} = -\frac{(D-2)}{m^2L^2} \g_{\mu\nu} \,.\eea
By plugging this relation into the generalized Gibbons-Hawking terms, we obtain the effective Gibbons-Hawking term which is given by the usual Gibbons-Hawking term of the Einstein gravity with rescaled Newton's constant as
\bea
 S_{EGH} &=& -\frac{q}{\k^{2}} \int_{\partial M} d^{D-1} x \sqrt{-\g}\,  K \,. \label{baction2}
\eea

From the AdS/CFT correspondence, the boundary fields $ \g_{\m\n}, \hat{f}_{\mu\nu} $ act as sources of the corresponding operators in the dual CFT. As far as we can consistently turn off the ghost-like massive modes by  boundary conditions, the form of the effective actions in the bulk and the boundary, in (\ref{action2}) and (\ref{baction2}), respectively, become exactly those of Einstein gravity. As a result, the effective boundary action for noncritical gravity becomes exactly the one for Einstein gravity up to an overall rescaling. This means that, as far as we remain at the tree level,  noncritical  gravity  reduces to Einstein gravity on AdS spacetime and thus the dual CFT of noncritical gravity should be identical to the one of pure Einstein gravity in the large $N$ limit. In the rest of the paper, we study various aspects of the dual CFT of noncritical  gravity and confirm this statement.

\section{The Fluid/Gravity Correspondence for Noncritical Gravity }
Fluid dynamics describes collective motions of huge number of constituent particles as continuum, of which construction is a predecessor of various modern field theories. Basic assumption in this coarse-grained description is that each small portion of fluid behaves as an entity. In other words, the length scale in the system is much larger than the `mean free path' of constituent particles, so the system is well approximated by continuum field variables.
The relevant variables to describe fluid dynamics are slowly varying local temperature $T(x)$ and fluid velocity $u^\mu (x)$. In the case of the neutral fluid,  fluid dynamics is governed by the conservation of energy-momentum tensor.   The main theme of  fluid dynamics is to determine the energy-momentum tensor as derivative expansions of $T(x)$ and $u^\mu(x)$ and find  coefficients of dissipation terms, so-called, transport coefficients.

A universal feature of fluid dynamics tells us that CFT also has a regime described by fluid dynamics. On top of this,  CFT has a holographic dual description by   gravity (or string theory)  on AdS space according to the AdS/CFT correspondence.  As a result, it is  intriguing and natural problem to construct fluid dynamics holographically, which was achieved in \cite{Policastro:2001yc}\cite{Bhattacharyya:2008jc}  and afterwards named as the fluid/gravity correspondence.

In this section we use the fluid/gravity correspondence to determine transport coefficients of the dual fluid of noncritical gravity. As will be shown later in this section, all the transport coefficients for noncritical gravity turn out to be the same as those for Einstein gravity, up to an overall factor.

\subsection{Fluid dynamics and dual geometry}
 In this subsection, we summarize some basics on fluid dynamics, partly to set up the notations.
 The usual formulation of fluid dynamics consists of two elements. One is the Navier-Stokes equation, which is based on the Newton's  second law of motion, and the other is  the continuity equation, which is just the macroscopic incarnation of microscopic conservation law. In the relativistic setup, these two equations are simply represented by a single conservation equation of the energy-momentum tensor as
\beq \nabla_{\mu}T^{\mu\nu} =0\,,\eeq
which is the main dynamical equation in the following.

The construction of fluid dynamics reveals its statistical nature and  so  it appears as the low energy description of any quantum field theory at sufficiently long wavelength limit. Because of such statistical nature,  relevant variables in relativistic fluids are  local thermodynamic variables: temperature $T(x)$ and velocity $u^\m(x)$.     See \cite{Rangamani:2009xk} for the review and details. 

Now we summarize some important formulae to fix our conventions for $d$-dimensional neutral fluids. Denoting density and pressure as $\rho(x)$ and $P(x)$, respectively, the energy-momentum tensor of  fluids can be written as the derivative expansion:
\beq T^{\mu\nu} = \Big[P(x) + \rho(x)\Big]u^{\mu}(x)u^{\nu}(x) + P(x)\gamma^{\mu\nu} + \Pi^{\mu\nu}(x)\,, \eeq
where $\Pi^{\mu\nu}=\sum_{n=1}^\infty \Pi^{\mu\nu}_{(n)}$ represents dissipation effects and the contribution of $n$-derivatives of $T(x)$ and $u^{\mu}(x)$.   

We choose  the, so-called, Landau frame, in which  the dissipation part is transverse to $u^{\mu}$,
\[ \Pi^{\mu\nu}u_{\nu}=0\,. \]
It is convenient to introduce the projection operator $P^{\m\n}= \gamma^{\mu\nu} + u^{\mu}u^{\nu}$.
The first order term $\Pi^{\mu\nu}_{(1)}$ can be decomposed as the traceless  and the trace part as
\beq
\Pi^{\mu\nu}_{(1)} = -2\eta\, \sigma^{\mu\nu} -\zeta \theta\, P^{\mu\nu}\,,  \eeq
where $\eta$ denotes shear viscosity, $\zeta$ does  bulk viscosity and 
\bea       \theta  &\equiv & P^{\mu\nu}\nabla_{\mu}u_{\nu}\,,   \nn  \\
        \sigma^{\mu\nu}  &\equiv& P^{\mu\alpha}P^{\nu\beta}\nabla_{(\alpha}u_{\beta)} - \frac{1}{d-1}\theta\, P^{\mu\nu}\,. \nn
\eea

In the context of the fluid/gravity correspondence, relevant fluids are  conformal ones which have additional constraints by underlying conformal symmetry.
For example, the bulk viscosity vanishes, $\zeta=0$, in the case of conformal fluids. In the following we consider conformal fluids  to study the fluid/gravity correspondence for noncritical gravity.

In the Weyl covariant formalism, the second order dissipation part for conformal fluids  consists of the following five terms: 
 \bea
\CI^{\mu\nu}_{1} &\equiv& 2u^{\alpha}D_{\alpha}\sigma^{\mu\nu} = 2\bigg[u^{\alpha}\nabla_{\alpha}\sigma^{\mu\nu} + \frac{1}{d-1}\sigma^{\mu\nu}\, \nabla^{\alpha}u_{\alpha} -  2u^{(\mu}\sigma^{\nu)\beta}\, u^{\alpha}\nabla_{\alpha}u_{\beta} \bigg]\,,   \\
 \CI^{\mu\nu}_{2} &\equiv& C^{\mu\alpha\nu\beta}u_{\alpha}u_{\beta}\,,   \qquad  \qquad
\CI^{\mu\nu}_{3}  \equiv 4\sigma^{\alpha \langle\mu}\sigma^{\nu \rangle}_{~~\alpha}\,, \nn \\
  \CI^{\mu\nu}_{4} &\equiv& 2\sigma^{\alpha \langle \mu}\omega^{\nu \rangle}_{~~\alpha}\,, ~~\qquad  \qquad  \CI^{\mu\nu}_{5} \equiv\omega^{\alpha \langle\mu}\omega^{\nu\rangle}_{~~\alpha}\,, \nn
\eea
where the bracket $\langle~\rangle $ around the indices makes a tensor  traceless and   transverse to $u^{\mu}$, {\it i.e.}  $ A^{\langle\mu\nu\rangle} \equiv P^{\mu\alpha}P^{\nu\beta}A_{(\alpha\beta)} - \frac{1}{d-1}P^{\mu\nu}P^{\alpha\beta}A_{\alpha\beta} $ and 
$\omega^{\mu\nu} = P^{\mu\alpha}P^{\nu\beta}\nabla_{[\beta}u_{\alpha]}$ denotes the vorticity tensor.
The second order dissipation part of  conformal fluids is given by
\bea
 \Pi^{\mu\nu}_{(2)} = \tau_{\pi}\eta\, \CI^{\mu\nu}_{1}  + \kappa\, \CI^{\mu\nu}_{2}  +\lambda_1\, \CI^{\mu\nu}_{3} + \lambda_2\, \CI^{\mu\nu}_{4}  +\lambda_3\, \CI^{\mu\nu}_{5}\,,
\eea
where $\tau_{\pi}$, $\kappa$ and $\lambda_{i}$ are called as second order  transport coefficients.

Since fluid dynamics describes a thermodynamic system by construction, the holographic dual geometry should be a finite temperature system, like black holes/branes. It turns out that the relevant dual geometry is a  planar-type black brane. Moreover,  the temperature  throughout fluids is  not a constant but the local, slow-varying, function of the position, which represents the nature of fluids at local thermal equilibrium. This  is realized in the gravity side as the local position dependence of black brane horizon.

One of the novel aspects of $D$-dimensional noncritical  gravity is the existence of the planar black brane solution,
\beq ds^2 = L^2\left[\frac{dr^2}{r^2 f(br)}  - r^2 f(br)dt^2 + r^2 d{\bf x}_{D-2}^2\right]\,, \label{blacksol}\eeq
where
\[
 f(r) = 1- \frac{1}{r^{D-1}}\,.
\]
One can easily confirm that the metric $g_{MN}$ in (\ref{blacksol}) satisfies the equation of motion (\ref{4eom}) by noting that the corresponding geometry is an Einstein manifold where the Ricci tensor  is given by
\beq
R_{MN} = -\frac{(D-1)}{L^2} g_{MN} \,. \eeq

The Hawking temperature of the planar black brane is proportional to the horizon radius and is given by
$
T = \frac{D-1}{4\pi b}\frac{1}{L} \,.
$
In what follows, we use this black brane solution to study the dual fluid dynamics of  noncritical  gravity.
The Bekenstein-Hawking entropy density $s$ of the planar black brane can be obtained by using the Wald formula with the original action $S$ in (\ref{action1}) and is given by
\bea
s = - 2\pi \sqrt{\g} \left. \frac{\p \CL}{\p R_{abcd}} \epsilon_{ab} \epsilon_{cd}\right|_{\rm horizon}
  = \frac{2\pi }{\k^2} \left (\frac{ 1}{ b}\right)^{D-2}\s \left[1 - \frac{(D-2)}{2 m^2 L^2} \right]  \,. \label{entropy}
\eea
One may note that the  same result can be obtained directly from  the effective action $S_{eff}$ in (\ref{action2}),
\beq
s =- 2\pi \sqrt{\g} \left. \frac{\p \CL_{eff}}{\p R_{abcd}} \epsilon_{ab} \epsilon_{cd}\right|_{\rm horizon}=\frac{2\pi q}{\k^2} \left (\frac{ 1}{ b}\right)^{D-2}\,,
\eeq
where $q$ is given by (\ref{q}). This gives an evidence of the nonlinear, classical equivalence between noncritical gravity and Einstein gravity on this AdS planar black brane background.


One may boost, with boost parameter $\beta_i$, the AdS planar  black brane solution along the translationally invariant spatial coordinates $x^i$ and write the metric using ingoing Eddington-Finkelstein coordinate as 
\beq
ds^2 =  L^2\left[-2 u_\m dx^\m dr - r^2 f(br) u_\m u_\n dx^\m dx^\n + r^2 P_{\m\n} dx^\m dx^\n \right]\,, \label{boosted}
\eeq
where
\beq
 u^\m = \Big(  \frac{1}{\sqrt{1 - \b^2}}, \frac{\b_i}{\sqrt{1-\b^2}} \Big)\,,
\eeq
may correspond to the globally uniform fluid velocity and $P_{\m\n} = \eta_{\m\n} + u_\m u_\n$.
Starting from this boosted black brane metric, we may implement slowly varying fluid velocity and temperature  by giving the position dependence  as $u_\m=u_\m(x)$ and $T=T(x)$. Accordingly, we consider the metric,
\beq ds^2 = L^2\left[-2u_\mu(x)dx^{\mu}dr -r^2f(b(x)r)~  u_{\mu}(x)u_{\nu}(x)dx^{\mu}dx^{\nu} + r^2 P_{\mu\nu}(x)dx^{\mu}dx^{\nu}\right]\,, \label{perturbed}\eeq
which  is, surely, not  the solution of the equations of motion for arbitrary $\b_i(x)$ and $b(x)$. To satisfy  the equations of motion we need to add correction terms in $ g_{MN}$, $\b_i(x)$ and $b(x)$ which  can be done by  systematic perturbation. Once the gravity solution is obtained perturbatively, one can determine the boundary energy-momentum tensor as will be described in the next section.

\subsection{The Fluid/Gravity Correspondence with Higher Curvature Terms }
Now we turn to the holographic description of the $d$-dimensional fluid dynamics corresponding to noncritical gravity in $D=d+1$ dimensions.
The fluid/gravity correspondence gives us the computational tool for  transport coefficients in the strong coupling regime.  According to the standard AdS/CFT prescription\cite{Balasubramanian:1999re},   the energy-momentum tensor of the $d$-dimensional CFT is identified with Brown-York tensor\cite{Brown:1992br}, up to counter terms, for the ADM-decomposed metric, (\ref{ADM}), of $D$-dimensional asymptotically AdS space\footnote{See~\cite{Kwon:2011jz} for some subtle issues on counter terms in higher derivative gravity.} .

After plugging the bulk equations of motion in  the bulk action, the leftover boundary terms combine with the Gibbons-Hawking boundary terms   and become the boundary action $S_{B}$. The Brown-York tensor is given by the functional derivative of the  boundary action $S_{B}$ with respect to the boundary metric $\g_{\m\n}$,
\begin{equation}
 T^{\m\n}_{BY} = -\frac{2}{\sqrt{-\gamma}} \frac{\delta S_{B}}{\delta \g_{\m\n}}~.
\end{equation}
The Brown-York tensor for the bulk action (\ref{action11}) and the generalized Gibbons-Hawking terms (\ref{baction1}) is found to be
\bea
8\pi G\,  T^{\m\n}_{BY} &=& \s\left[1 - \frac{1}{2(D-2)} (\hat{s}-\hat{f})\right] ( K^{\m\n} - K \g^{\m\n})  \label{BY}\\
&& + \frac{\s}{(D-2)}\left[ \nabla^{(\m}\hat{h}^{\n)} - \half D_{r} \hat{f}^{\m\n} + K_{\r}^{~(\m}\hat{f}^{\n) \r}- \gamma^{\m\n} \Big( \nabla_\r  \hat{h}^\r
    - \half D_r\hat{f} \Big) \right]\,, \nn\eea
where hatted quantities are defined by (\ref{hatfuc}) and
\beq  \hat{s}=N^2 s\,, \quad   \hat{h}^\m= N (h^\m+sN^\m)\,,  \label{hatfuc1} \eeq
and the covariant
derivatives along $r$ are
\bea
D_r \hat{f}^{\m\n} &=& \frac{1}{N}\Big(\partial_r\hat{f}^{\m\n} -N^\r \p_\r \hat{f}^{\m\n} + \hat{f}^{\r\n}\p_\r N^\m + \hat{f}^{\m\r}\p_\r N^\n\Big)\,, \nn \\
D_{r}\hat{f} &=& \frac{1}{N}\Big(\p_{r}\hat{f} -N^\m\p_\m\hat{f}\Big)\,.   \eea

We find that the perturbative solutions starting from the metric (\ref{perturbed}) of noncritical gravity has exactly the same form as those of Einstein gravity. The perturbative solution satisfies
\beq
R_{MN} = -\frac{(D-1)}{L^2} g_{MN} \,, \qquad f_{MN} = -\frac{(D-2)}{m^2L^2}g_{MN} \,.
\eeq
 Inserting this solution into the expression of  Brown-York tensor in (\ref{BY}), one can read off various transport coefficients holographically. 
By using (\ref{aux1}), we can find the hatted quantities in (\ref{BY}):
\bea \hat{s} &=& - \frac{(D-2)}{m^2 L^2}\,,\qquad \hat{h}^{\m} = 0 \,,\nn\\
     \hat{f}^{\m\n} &=& - \frac{(D-2)}{m^2 L^2} \g^{\m\n} \,,\qquad \hat{f} = - \frac{(D-1)(D-2)}{m^2 L^2}\,, \\
     D_r \hat{f}^{\m\n} &=& - \frac{2(D-2)}{m^2 L^2} K^{\m\n} \,,\qquad D_r \hat{f} = 0 \,. \nn
\eea
Then the Brown-York tensor (\ref{BY}) becomes
\beq
8\pi G T_{BY}^{\m\n} = \s \left[ 1 - \frac{(D-2)}{2m^2 L^2} \right]  (K^{\m\n} - K \g^{\m\n}) = q (K^{\m\n} - K \g^{\m\n})\,,
\eeq
which is identical with the one obtained directly from the effective action (\ref{action2}) and  the effective Gibbons-Hawking term  (\ref{baction2}).
Therefore we find 
\bea
  T^{\m\n}_{BY} ({\rm noncritical}) &=& q~ T^ {\m\n}_{BY}({\rm Einstein})~,\label{EMtensor}\eea
which  confirms the nonlinear, classical  equivalence between noncritical gravity and Einstein gravity.

From the above  one can readily read off the transport coefficients in the dual fluid of noncritical gravity from those corresponding to  Einstein gravity. The viscosity  is given by
\beq
\eta = q\frac{L^{d-1}}{2\k^2} \left(\frac{4\pi}{d}\, T \right)^{d-1} =\frac{q}{2\k^2}\frac{1}{b^{d-1}}\,,  \eeq
and the viscosity-to-entropy ratio saturates the KSS bound:
\beq
\frac{\eta}{s} = \frac{1}{4\pi}\,.
\eeq
We stress that this is an exact result, all orders in $1/m^2$.  

The transport coefficients at the second order  are given by
\begin{eqnarray}
\tau_\pi =  \frac{d}{4\pi T}\left[1+\frac{1}{d}\,{\rm Harmonic}\Big(\frac{2}{d}-1\Big)\right]\,, \qquad \kappa=\frac{d}{2\pi(d-2)}\frac{\eta}{T}\,, \\
\lambda_1=\frac{d}{8\pi }\frac{\eta}{T}\,, \qquad \lambda_2=\frac{1}{2\pi }\,{\rm Harmonic}\Big(\frac{2}{d}-1\Big)\,\frac{\eta}{T}\,, \qquad \lambda_3=0\,, \nn
\end{eqnarray}
where ${\rm Harmonic}(x)$ is harmonic number function.

These results are novel as they are all order expressions in $1/m^2$.  Note that $m^2$ enters nonlinearly in the expression of  $q$, (\ref{q}), through $L^2(m)$ given  in  (\ref{NewCos}). After expanded in terms of $1/m^2$,  our results agree with the known perturbative expressions of transport coefficients in the dual fluid of our class of higher curvature gravity. Explicitly, the shear viscosity from noncritical gravity, after expanded in terms of $1/m^2$, is given  by
\beq
\eta =    \frac{1}{2\k^2} \frac{1}{b^{d-1}} \, \sigma\left[ 1 - \frac{d-1}{2m^2\ell^2} + \CO\Big(\frac{1}{m^4}\Big)\right]\,.
\eeq

As was emphasized in the previous sections, our results go beyond those from field redefinition  or other perturbative approaches like the effective action method \cite{Kats:2007mq}\cite{Banerjee:2009wg}.
Though the higher curvature terms in noncritical gravity can be absorbed into the Einstein-Hilbert term via a field redefinition up to $1/m^2$ order,  this field redefinition obscures higher order contribution in $1/m^2$.  The effective action method, on the other hand,  may address higher orders systematically. However, nonlinear transport coefficients, $\lambda_i (i=1,2,3)$ are out of reach in  this effective action method.  In contrast we found the exact expressions for the transport coefficients in the dual fluid dynamics of non critical gravity. 

These holographic transport coefficients
are formally related to the boundary $n$-point correlation functions in Euclidean space as
\begin{eqnarray}
\langle T^{\m\n}(0)\rangle &=&  \langle T^{\m\n}(0)\rangle_{0}+\frac{1}{2}\int d^{d}x  \langle T^{\m\n}(0)T^{\alpha\beta}(x)\rangle_{0}h_{\alpha\beta} \\
&&+\frac{1}{8}\int d^{d}x\, d^{d}y \,\langle T^{\m\n}(0)T^{\alpha\beta}(x)T^{\g\delta}(y)\rangle_{0}h_{\alpha\beta}(x)h_{\g\delta}(y)+\cdots~. \nn
\end{eqnarray}
For example, the Kubo formula in linear response theory in Minkowski space is given by
\beq
\eta = \lim_{\omega\rightarrow 0} \frac{1}{2\omega} \int dt d{\vec{x}} e^{i\omega t} \langle [ T_{12}(x), T_{12}(0) ] \rangle_0 =  - \lim_{\omega \rightarrow 0 } \frac{1}{\omega} {\rm Im}\,G^{R}(\omega, 0) \,,
\eeq
where $G^R$ is the retarded Green's function. The second order transport coefficients are generically related to the three-point correlation functions. (See, for example, \cite{Moore:2010bu} for more details.) As mentioned in the previous section,  the $n$-point correlation functions in noncritical gravity should be identical to those from Einstein gravity up to rescaling of Newton's constant and the cosmological constant. What we found is that the our result (\ref{EMtensor}) is consistent with  all the nonlinear transport coefficients  from  the $n$-point correlation functions in noncritical gravity.

\section{Holographic Entanglement Entropy}

Entanglement entropy is one of the most interesting measure how two systems, $\CA$ and its complement,  are quantum-mechanically related. When two systems are entangled quantum-mechanically, the measurement of one system leads to some correlated information on the other system through quantum superposition.  Formally, the entanglement entropy of $\CA$   is defined by the partial trace  of von Neumann entropy  over the complement of $\CA$. It measures how much system $\CA$ is entangled with its complement. 
(See \cite{Calabrese:2009qy} for some review.) 
Since  the AdS/CFT correspondence may be regarded as a quantum principle, it may realize this entanglement entropy holographically, which is indeed done in~\cite{Ryu:2006bv} and coined as holographic entanglement entropy(HEE). According to this proposal in Einstein gravity, the entanglement entropy of region $\CA$ in the CFT side corresponds to the  minimal area of codimension two hypersurface $\Sigma_\CA$ in AdS space homologous to the domain $\CA$ at the asymptotic boundary\cite{Fursaev:2006ih}\cite{Headrick:2010zt}.  This proposal is extended to a certain higher curvature gravity, {\it i.e.} Lovelock gravity,   in~\cite{deBoer:2011wk}.

To compute the HEE for noncritical gravity whose action contains a specific combination of the higher curvature terms, one should extend or adjust the previous proposals. There is no known universal prescription to obtain the HEE for general higher derivative gravity.  Luckily, the extension is rather simple because of the nature of noncritical gravity.  It was shown in \cite{deBoer:2011wk} that the Wald formula applied on $\Sigma_\CA$ for the HEE does not give the  right result in the case of the Lovelock gravity. In contrast we show that  the Wald formula for the HEE gives the correct answer in the case of noncritical gravity. See~\cite{Kwon:2012tp} for the three-dimensional case.

In this section, we confirm that the HEE from the effective action (\ref{baction2}) is identical with the HEE of noncritical gravity. This supports our claim  on the nonlinear equivalence between Einstein gravity and noncritical gravity at the tree level. In order to establish the classical equivalence  we obtain the HEE for noncritical gravity using  two independent but related methods in  the following. Firstly,  we use the Wald  formula on the AdS background and then we use the replica method on the Euclidean AdS soliton background. We find that both of them satisfy basic properties of the HEE as much as those from Einstein gravity.
 We also comment on R\'enyi entropy.

Let us consider the Wald  formula on $\Sigma_\CA$ for the HEE in the case of the AdS background. In this approach, we take into account the, so-called, central charge function which appears in holographic c-theorem. This function is monotonic under the renormalization group  flow and coincides with a certain central charge at the conformal points. To introduce this function, let us consider the following geometry which corresponds to the renormalization group flow in the CFT side,
\beq ds^2 = L^2\bigg[dr^2 + e^{2A(r)}\Big(-dt^2 + d{\bf x}^2_{d-1}\Big)\bigg]\,, \eeq
where AdS spacetime is represented by $A(r)=r$.

Borrowing the results from~\cite{Myers:2010tj}, one can see that the central charge function of $d$-dimensional dual CFT of noncritical gravity is given  by
\beq a_{d}(r) = \frac{\sigma\pi^{d/2}}{\Gamma\big(\frac{d}{2}\big)(A'(r))^{d-1}} \frac{L^{d-1}}{\kappa^2} \bigg[1 - \frac{d-1}{2m^2L^2}A'^2(r)\bigg]\,. \eeq
We propose that the HEE, $\CS_\CA$, for the region $\CA$ can be obtained by using the Wald formula in noncritical gravity. If $\CS_W$ denotes the `entropy' expression by the Wald formula  for the hypersurface $\Sigma_\CA$ as

\beq
\CS_W = - 2\pi\int_{\Sigma_\CA} d^{d-1}x \sqrt{\gamma_E}  \frac{\p \CL}{\p R_{abcd}} \epsilon_{ab} \epsilon_{cd}\bigg|_{AdS}\,,
\eeq
our proposal simply means  $\CS_\CA = \CS_W$ for noncritical gravity.

It was shown in \cite{Myers:2010tj} that, in general,  $\CS_{W}$  can be expressed in terms of  $a^{*}_{d} \equiv a_{d}(r)|_{AdS}$ as
\beq \CS_{W} =  \frac{2\pi}{\pi^{d/2}}\frac{\Gamma\big(\frac{d}{2}\big)}{L^{d-1} } ~ a^{*}_{d} \int_{\Sigma_\CA} d^{d-1}x\sqrt{\gamma_E}\,. \eeq
  In even $d$-dimensions $a^{*}_{d}$ is nothing but the central charge for the A-type trace anomaly and in odd dimensions it corresponds to the renormalized partition function of dual CFT.

Now, after plugging $a^{*}_{d} =q\frac{\pi^{d/2}}{\Gamma\big(\frac{d}{2}\big)} \frac{L^{d-1}}{\kappa^2}$ in $\CS_W$, one can see that  the HEE of region $\CA$ in noncritical gravity is given by
\beq \CS_{\CA} =\CS_W= q\frac{2\pi}{\kappa^2}{\rm Area}(\Sigma_\CA)\,. \eeq
This is identical with the HEE from the effective action (\ref{baction2}). This  also confirm the equivalence between noncritical gravity and Einstein gravity at low energy.

Secondly, we adopt the replica method on the D-dimensional Euclidean AdS soliton which may be written as
\beq
ds^2 =L^2 \bigg[ \frac{ dr^2}{r^2 f(r)} + r^2 f(r) d\theta^2 + r^2 (dt_E^2 + d{\bf{x}}_{d-2}^2)\bigg]\,,
\eeq
where
\beq f(r) = 1 - \left(\frac{r_0}{r}\right)^{d}\,,
\eeq
and $\theta$ is periodic, $\theta\sim \theta +  \frac{4\pi }{d \, r_0}$, so that the near horizon geometry is regular.  

Following  \cite{Ogawa:2011fw} we consider the region $\CA$ given by half space $x_1>0$ at $t_E=0$.  In order to compute the HEE $\CS_\CA$ via the replica method, we introduce a conical singularity and  place a deficit angle $\d = 2\pi (1-n)$ on the $(d-1)$-dimensional surface $\S_\CA$ defined by $x_1=t_E =0$. Then the curvature tensors behave like \cite{Fursaev:1995ef}
\beq
 \CR_{MN} = R_{MN} + 2\pi (1-n)\left(n^{t_E}_M n^{t_E}_N  + n^{x_1}_M n^{x_1}_N\right) \d(\S_\CA)\,,\quad
 \CR = R + 4\pi (1-n) \d(\S_\CA)\,,
\eeq
where $ R_{MN}$ and $R$ denote the Ricci tensor and Ricci scalar without singularities while
 two orthonormal vectors $n^{t_E}_M$ and $n^{x_1}_M$ are orthogonal to the surface $\S_\CA$.

The HEE through the the replica method is given by
\beq
\CS_{\CA} = - \frac{\p}{\p n} \log \left. \frac{Z_n}{(Z_1)^n}\right|_{n=1} \simeq \left. \frac{\p I_n}{\p n}\right|_{n=1} - I_1\,,
\eeq
where $I_n$ is the on-shell Euclidean action value for $n$-sheeted AdS soliton with conical singularity of deficit angle $2\pi (1-n)$. As noted in \cite{Ogawa:2011fw} appropriate surface terms should be included in the action to cancel the divergent part of the entanglement entropy. We have omitted these terms since these are not so relevant in our discussion. 

The contribution to  the HEE from the Einstein-Hilbert and the cosmological constant terms is given by
\beq
\CS_{\CA}^{(EH)}= \frac{2\pi\s}{\k^2}{\rm{Area}}(\Sigma_\CA)\,.
\eeq
The contribution from the higher curvature terms in noncritical gravity can be computed  as
\beq
\CS_{\CA}^{(R^2)} =\left. \frac{\p I^{(R^2)}_n}{\p n}\right|_{n=1} - I^{(R^2)}_1  = - \frac{\pi\s ( d-1)}{\k^2 m^2 L^2} {\rm{Area}}(\Sigma_\CA) \,.
\eeq
Therefore, total HEE becomes
\bea
\CS_{\CA} = S_{\CA}^{(EH)} + S_{\CA}^{(R^2)}
      = q \frac{2\pi  }{\k^2} {\rm{Area}}(\Sigma_\CA)\,,
\eea
where $q$ is given by (\ref{q}).
Once again, this shows the nonlinear, classical  equivalence between noncritical gravity and Einstein gravity.

Another interesting quantity  related to the entanglement entropy  is the, so-called, R\'{e}nyi entropy which  is more refined entropy than the entanglement entropy. The R\'{e}nyi entropy is defined by
\beq \CS^{n}_{\CA} \equiv \frac{1}{1-n}\ln \tr\, \big[\rho^{n}_{\CA}\big]\,, \eeq
where $n$ is a positive real number.   Note that this quantity becomes entanglement entropy  when $n\rightarrow 1$.   Interestingly, R\'{e}nyi entropy for spherically entangled domain in the CFT side can be described by thermal entropy   and then by black hole geometry in the dual gravity side~\cite{Hung:2011nu}. Concretely, the relevant geometry for the R\'{e}nyi entropy  computation on the spherical domain  turns out to be  topological black holes whose metric is given in the form of
\beq
ds^2 = - \Big( \frac{r^2}{L^2} f(r) - 1 \Big) dt^2 + \frac{dr^2}{\frac{r^2}{L^2} f(r) - 1} + r^2 dH_{d-1}^2\,,
\eeq
where $dH_{d-1}^2$ denotes the metric for the hyperbolic space with unit radius.
In noncritical gravity, we find that  the function $f(r)$ takes the same form as the one  in Einstein gravity:
\beq
f(r) = 1 - \frac{c}{r^d}\,,      \eeq
where the constant $c$ is related to  the position of the horizon, $r_H$  as $c = r^d_H - L^2r^{d-2}_H$. It is straightforward to compute the   thermal entropy    by the Wald formula. 
Through the relation between R\'{e}nyi and thermal entropy,   
one can check that holographic R\'{e}nyi entropy in noncritical gravity is given by
\beq \CS^{n}_{\CA}\,  ({\rm noncritical})  = q~ \CS^{n}_{\CA}\, ({\rm Einstein})\,, \eeq
where $\CA$ is a spherical domain.

\section{Conclusion}
In this paper we studied a class of higher derivative gravity in arbitrary dimensions, named as noncritical gravity, which is defined on asymptotically AdS spacetime. We 
found that noncritical gravity is classically equivalent to Einstein gravity on the same type of background at the full nonlinear level.  The equivalence implies that the dual CFT of noncritical gravity is the same as the one of Einstein gravity in the large $N$ limit. 
In particular, $n$-point correlation functions of the energy-momentum tensor in the CFT corresponding to noncritical gravity should be the same as those from Einstein gravity. 

From the side of noncritical gravity, we calculated the transport coefficients in the dual fluid  and obtained the same form as those from Einstein gravity at the nonlinear level. As a byproduct we found the viscosity-to-entropy ratio which saturates the KSS bound to all orders in the coupling of higher derivative terms. We also found the second order transport coefficients to all orders in the coupling.

Furthermore we studied the HEE for noncritical gravity. We found that  in this case  the Wald formula can be used to compute the HEE. We also used the replica method and obtained consistent results on the HEE. Once again we found the results which agree with those from Einstein gravity.

All these results support our claim on the equivalence between noncritical gravity and Einstein gravity. We have shown that noncritical gravity is well-defined at low energy and can be used to study the boundary CFT. 

It would be interesting to generalize four-dimensional $\CN=1$ NEW supergravity to higher dimensions which may correspond to the supersymmetrization of noncritical gravity. It would be also interesting to study gravity whose action contains even higher curvature terms of Ricci scalar and Ricci tensor only. 

\section*{Acknowledgments}
SHY would like to thank S. Nam, J.D.  Park and Y. Kwon at Kyunghee Univ. for useful discussion.
SH is supported in part by the National Research Foundation of Korea(NRF)  grant funded by the Korea government(MEST) with  the grant number  2009-0074518 and the grant number 2009-0085995. SH, WJ and SHY are supported by the grant number 2005-0049409 through the Center for Quantum Spacetime(CQUeST) of Sogang University. JJ is supported by the National Research Foundation of Korea(NRF)  grant funded by the Korea government(MEST) with  the grant number 2009-0072755.


\newpage


\begin{thebibliography}{999}

\bibitem{Maldacena:2011mk}
  J.~Maldacena,
  ``Einstein Gravity from Conformal Gravity,''
 [arXiv:1105.5632 [hep-th]].

\bibitem{Lu:2011ks}
  H.~Lu, Y.~Pang and C.~N.~Pope,
  ``Conformal Gravity and Extensions of Critical Gravity,''
  Phys.\ Rev.\  D {\bf 84} (2011) 064001
  [arXiv:1106.4657 [hep-th]].

\bibitem{Lu:2011zk}
  H.~Lu and C.~N.~Pope,
  ``Critical Gravity in Four Dimensions,''
  Phys.\ Rev.\ Lett.\  {\bf 106} (2011) 181302
  [arXiv:1101.1971 [hep-th]].

\bibitem{Lu:2011mw}
  H.~Lu, C.~N.~Pope, E.~Sezgin and L.~Wulff,
  ``Critical and Non-Critical Einstein-Weyl Supergravity,''
  [arXiv:1107.2480 [hep-th]].

\bibitem{Hyun:2011ej}
  S.~Hyun, W.~Jang, J.~Jeong and S.~-H.~Yi,
  ``Noncritical Einstein-Weyl Gravity and the AdS/CFT Correspondence,''
  JHEP {\bf 1201} (2012) 054
  [arXiv:1111.1175 [hep-th]].

\bibitem{Deser:2011xc}
  S.~Deser, H.~Liu, H.~Lu, C.~N.~Pope, T.~C.~Sisman and B.~Tekin,
  ``Critical Points of D-Dimensional Extended Gravities,''
  Phys.\ Rev.\  D {\bf 83} (2011) 061502
  [arXiv:1101.4009 [hep-th]].

\bibitem{Kovtun:2003wp}
  P.~Kovtun, D.~T.~Son and A.~O.~Starinets,
  ``Holography and hydrodynamics: Diffusion on stretched horizons,''
  JHEP {\bf 0310} (2003) 064
  [hep-th/0309213].

\bibitem{Kovtun:2004de}
  P.~Kovtun, D.~T.~Son and A.~O.~Starinets,
  ``Viscosity in strongly interacting quantum field theories from black hole physics,''
  Phys.\ Rev.\ Lett.\  {\bf 94} (2005) 111601
  [hep-th/0405231].

\bibitem{Myers:2010tj}
  R.~C.~Myers and A.~Sinha,
  ``Holographic c-theorems in arbitrary dimensions,''
  JHEP {\bf 1101} (2011) 125
  [arXiv:1011.5819 [hep-th]];

  R.~C.~Myers and A.~Sinha,
  ``Seeing a c-theorem with holography,''
  Phys.\ Rev.\ D {\bf 82} (2010) 046006
  [arXiv:1006.1263 [hep-th]].
  
\bibitem{arXiv:1102.4091}
  E.~A.~Bergshoeff, O.~Hohm, J.~Rosseel and P.~K.~Townsend,
  ``Modes of Log Gravity,''  Phys.\ Rev.\ D\ {\bf 83} (2011) 104038  [arXiv:1102.4091 [hep-th]].  

\bibitem{Breitenlohner:1982jf}
  P.~Breitenlohner and D.~Z.~Freedman,
  ``Stability in Gauged Extended Supergravity,''
  Annals Phys.\  {\bf 144 } (1982)  249.


\bibitem{Bergshoeff:2009hq}
  E.~A.~Bergshoeff, O.~Hohm and P.~K.~Townsend,
  ``Massive Gravity in Three Dimensions,''
  Phys.\ Rev.\ Lett.\  {\bf 102} (2009) 201301
  [arXiv:0901.1766 [hep-th]];
      
  E.~A.~Bergshoeff, O.~Hohm, J.~Rosseel, E.~Sezgin and P.~K.~Townsend,
  ``More on Massive 3D Supergravity,''
  Class.\ Quant.\ Grav.\  {\bf 28} (2011) 015002
  [arXiv:1005.3952 [hep-th]].
   
\bibitem{Hohm:2010jc}
  O.~Hohm and E.~Tonni,
  ``A boundary stress tensor for higher-derivative gravity in AdS and Lifshitz
  backgrounds,''
  JHEP {\bf 1004}, 093 (2010)
  [arXiv:1001.3598 [hep-th]].

\bibitem{Policastro:2001yc}
  G.~Policastro, D.~T.~Son and A.~O.~Starinets,
   ``The shear viscosity of strongly coupled N = 4 supersymmetric Yang-Mills
  plasma,''
  Phys.\ Rev.\ Lett.\  {\bf 87}, 081601 (2001)
  [arXiv:hep-th/0104066].
  
\bibitem{Bhattacharyya:2008jc}
  S.~Bhattacharyya, V.~E Hubeny, S.~Minwalla and M.~Rangamani,
  ``Nonlinear Fluid Dynamics from Gravity,''  JHEP {\bf 0802} (2008) 045  [arXiv:0712.2456 [hep-th]].  
 
\bibitem{Rangamani:2009xk}
  M.~Rangamani,
  ``Gravity \& Hydrodynamics: Lectures on the fluid-gravity correspondence,''
  Class.\ Quant.\ Grav.\  {\bf 26} (2009) 224003
  [arXiv:0905.4352 [hep-th]];

  V.~E.~Hubeny, S.~Minwalla and M.~Rangamani,
  ``The fluid/gravity correspondence,''
  [arXiv:1107.5780 [hep-th]];
  
  N.~Banerjee and S.~Dutta,
  ``Holographic Hydrodynamics: Models and Methods,''
  [arXiv:1112.5345 [hep-th]].
  
\bibitem{Balasubramanian:1999re}
  V.~Balasubramanian and P.~Kraus,
  ``A stress tensor for anti-de Sitter gravity,''
  Commun.\ Math.\ Phys.\  {\bf 208} (1999) 413
  [[arXiv:hep-th/9902121]].

\bibitem{Brown:1992br}
  J.~D.~Brown and J.~W.~.~York,
  ``Quasilocal energy and conserved charges derived from the gravitational
  action,''
  Phys.\ Rev.\  D {\bf 47} (1993) 1407
  [arXiv:gr-qc/9209012].

\bibitem{Kwon:2011jz}
  Y.~Kwon, S.~Nam, J.~D.~Park and S.~H.~Yi,
  ``Holographic Renormalization and Stress Tensors in New Massive Gravity,''
  JHEP {\bf 1111} (2011) 029
  [arXiv:1106.4609 [hep-th]].

\bibitem{Kats:2007mq}
  Y.~Kats and P.~Petrov,
  ``Effect of curvature squared corrections in AdS on the viscosity of the dual gauge theory,''
  JHEP {\bf 0901} (2009) 044
  [arXiv:0712.0743 [hep-th]];
  
  M.~Brigante, H.~Liu, R.~C.~Myers, S.~Shenker and S.~Yaida,
  ``Viscosity Bound Violation in Higher Derivative Gravity,''
  Phys.\ Rev.\ D {\bf 77} (2008) 126006
  [arXiv:0712.0805 [hep-th]].
  
\bibitem{Banerjee:2009wg}
  N.~Banerjee and S.~Dutta,
  ``Higher Derivative Corrections to Shear Viscosity from Graviton's Effective Coupling,''
  JHEP {\bf 0903} (2009) 116
  [arXiv:0901.3848 [hep-th]].
  
  
  
\bibitem{Moore:2010bu}
  G.~D.~Moore and K.~A.~Sohrabi,
  ``Kubo Formulae for Second-Order Hydrodynamic Coefficients,''
  Phys.\ Rev.\ Lett.\  {\bf 106} (2011) 122302
  [arXiv:1007.5333 [hep-ph]].
  
 
  
\bibitem{Calabrese:2009qy}
  P.~Calabrese and J.~Cardy,
  ``Entanglement entropy and conformal field theory,''
  J.\ Phys.\ A  {\bf 42} (2009) 504005
  [arXiv:0905.4013 [cond-mat.stat-mech]].

 
  
\bibitem{Ryu:2006bv}
  S.~Ryu and T.~Takayanagi,
  ``Holographic derivation of entanglement entropy from AdS/CFT,''
  Phys.\ Rev.\ Lett.\  {\bf 96} (2006) 181602
  [hep-th/0603001];

  T.~Nishioka, S.~Ryu and T.~Takayanagi,
  ``Holographic Entanglement Entropy: An Overview,''
  J.\ Phys.\ A A {\bf 42} (2009) 504008
  [arXiv:0905.0932 [hep-th]].

\bibitem{Fursaev:2006ih}
  D.~V.~Fursaev,
  ``Proof of the holographic formula for entanglement entropy,''
  JHEP {\bf 0609}, 018 (2006)
  [arXiv:hep-th/0606184].
  
\bibitem{Headrick:2010zt}
  M.~Headrick,
  ``Entanglement Renyi entropies in holographic theories,''
  Phys.\ Rev.\  D {\bf 82} (2010) 126010
  [arXiv:1006.0047 [hep-th]].



\bibitem{deBoer:2011wk}
  J.~de Boer, M.~Kulaxizi and A.~Parnachev,
  ``Holographic Entanglement Entropy in Lovelock Gravities,''
  JHEP {\bf 1107} (2011) 109
  [arXiv:1101.5781 [hep-th]];

  L.~-Y.~Hung, R.~C.~Myers and M.~Smolkin,
  ``On Holographic Entanglement Entropy and Higher Curvature Gravity,''
  JHEP {\bf 1104} (2011) 025
  [arXiv:1101.5813 [hep-th]].
 
\bibitem{Kwon:2012tp}
  Y.~Kwon, S.~Nam, J.~D.~Park and S.~H.~Yi,
  ``AdS/BCFT Correspondence for Higher Curvature Gravity: An Example,''
  [arXiv:1201.1988 [hep-th]].
   
\bibitem{Ogawa:2011fw}
  N.~Ogawa and T.~Takayanagi,
  ``Higher Derivative Corrections to Holographic Entanglement Entropy for AdS Solitons,''  JHEP {\bf 1110} (2011) 147  [arXiv:1107.4363 [hep-th]].  
 
\bibitem{Fursaev:1995ef}
  D.~V.~Fursaev and S.~N.~Solodukhin,
  ``On The Description Of The Riemannian Geometry In The Presence Of Conical
  Defects,''
  Phys.\ Rev.\  D {\bf 52} (1995) 2133
  [arXiv:hep-th/9501127];

 
\bibitem{Hung:2011nu}
  L.~Y.~Hung, R.~C.~Myers, M.~Smolkin and A.~Yale,
  ``Holographic Calculations of Renyi Entropy,''
  JHEP {\bf 1112} (2011) 047
  [arXiv:1110.1084 [hep-th]].

 
    
  






\end{thebibliography}
\end{document}